\documentclass{article}

\usepackage[nonatbib,final]{neurips_2019}

\usepackage[numbers]{natbib}

\usepackage[utf8]{inputenc} 
\usepackage[T1]{fontenc}    
\usepackage{hyperref}       
\usepackage{url}            
\usepackage{booktabs}       
\usepackage{amsfonts}       
\usepackage{nicefrac}       
\usepackage{microtype}      

\title{Towards Sustainable Architecture: 3D Convolutional Neural Networks for Computational Fluid Dynamics Simulation and Reverse Design Workflow}

\author{
  Josef Musil*, Jakub Knir, Athanasios Vitsas, Irene Gallou \\
  Specialist Modelling Group\\
  Foster + Partners\\
  London, UK \\
  \texttt{*jmusil@fosterandpartners.com}
}

\usepackage{graphicx}
\graphicspath{{./}}

\begin{document}

\maketitle

\begin{abstract}

We present a general and flexible approximation model for near real-time prediction of steady turbulent flow in a 3D domain based on residual Convolutional Neural Networks (CNNs). This approach can provide immediate feedback for real-time iterations at the early stage of architectural design. This work-flow is then reversed and offers a designer a tool that generates building volumes based on target wind flow.

\end{abstract}

\section{Introduction}

Architectural design depends on environmental constraints from its initial stage, when buildings and cities take shape and informed decisions on sustainable development are especially important.
However, proposals can change rapidly and it is challenging to provide relevant simulations at the same pace. Especially, Computational Fluid Dynamics (CFD) requires complex geometry preparation and computationally demanding solutions. This is time consuming and in contradiction to the speed of design progress.
To facilitate the impact of CFD on design, this work focuses on data-driven flow field prediction and leverages the approximation using CNNs. 

Previous work shows good results in fast simulation of fluid dynamics \cite{fastCFD_Guo2016, fastcfd_fewSamples_White2019, wilkinson2014approximating} and approximation of Navier-Stokes equations \cite{navier_Miyanawala2017}. We focus on the application in architecture in 3D domains using CNNs with residual blocks \cite{residual_Tompson2017}. Our further contribution is to experiment with reverse training to predict architectural volumes.
Fast forward prediction has a significant potential to improve sustainable design, yet still to inform the design by the results of a CFD analysis depends on the creativity of the designer and there is no direct way to inform design decisions other than choosing the best design out of a high number of proposals. We facilitate this using the same CNN model trained in the reverse direction.

\section{Data Creation and Simulation}

Various geometries representing samples of urban structures were generated using a visual programming language and common Computer Aided Design (CAD) software. These samples consider standard distribution of building heights in a city, and width and depths were constrained to common minimum and maximum sizes. Each sample is represented as a 3D mesh and has to fit inside a volume with dimensions of 256m x 128m x 64m. Meshes are voxelised with 1m resolution. Our total set had 3500 samples: 3325 (95\%) for training and 175 (5\%) for testing.

In aerodynamics related design, analysis and optimisation problems, flow fields are simulated using CFD solvers. However, CFD simulation is usually a computationally expensive, memory demanding and time consuming iterative process. These drawbacks of CFD limit opportunities for design space exploration and interactive design.
Our data set was simulated using OpenFOAM software. Due to the number of cases needed for CNN training the workflow was fully automated.

\section{Neural Network Architecture}

Our network architecture has a U-net shape with eight encoder and seven decoder layers. Each layer consists of one residual block with one 3D convolution with stride 2 filters 4x4x4 and one 3D convolution with stride 1 and filter 3x3x3. Our tests showed these gated blocks improved our results compared to plain encoder decoder architecture \cite{encdec_ganNoDisc_img2img_Isola2017}. Activations are concatenated exponential linear units. Additionally, this fully connected CNN provides good generalisation properties for geometries different from the training set and for input data larger than the training sample dimensions. This network can estimate the wind velocity field three orders of magnitude faster than a CFD solver both in 3D domain. The test mean squared error loss kept improving throughout 1000 epochs for both directions and proved generalisation ability. In reverse direction we change the number of output channels to 1 (1 represents building, 0 outside space), compared to 3 output channels in forward direction (x, y, z components of wind direction vectors).

\section{Results}

\begin{figure}[t]
  \centering
  \includegraphics[width=\textwidth]{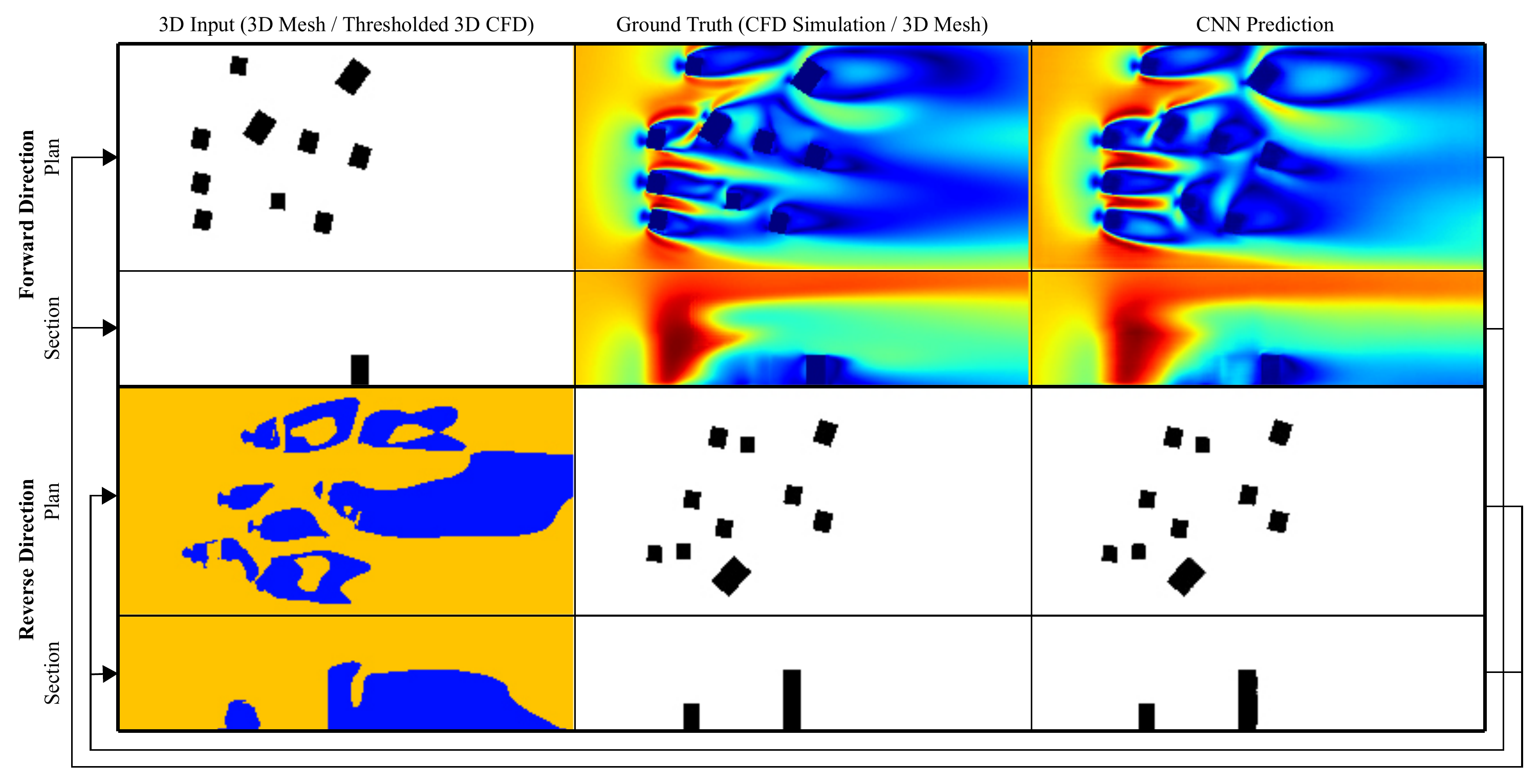}
  \caption{Forward (top) and reverse (bottom) prediction.}
  \label{fig-samples-both}
\end{figure}

We implemented a Flask server to interactively run prediction from a common CAD software Rhino, and its visual programming interface Grasshopper. This CAD provides visualization tools that were used to produce sample images. 
We show one sample of forward CFD prediction of the wind velocity magnitude (calculated as Froberius norm of x,y,z components) and one of reverse prediction of volumes in Figure \ref{fig-samples-both}. Yellow is undesirable high wind speed, blue is preferred low wind speed.

\section{Discussion}

In concept design stages, fast analysis responses are a critical objective across the industries. As near real time prediction proves valuable, the proposed methodology has potential applications outside the architecture.
In reverse order, it shifts designer's focus directly onto the desired effect and human well-being to use their time efficiently towards sustainable design.
In our further research, we are looking at improving the cost function by adding continuity equation error and implementing a generative adversarial network.
We also explore possibilities to generate multiple building predictions for a single input wind flow.

\newpage

\subsubsection*{Demo}
We present a short animated video that demonstrates this paper's work. The demo covers the whole design and simulation process as seen in the links below:
\begin{center}
  \url{https://youtu.be/GfJS8AI0omY} and \url{https://youtu.be/rKO4wei3Dos}
\end{center}

\subsubsection*{Ethical Implications}
This work does not involve major ethical issues. Only in the case of building a future general model, enough samples from variety of cultural and architectural styles and typologies should be considered.

\subsubsection*{Acknowledgments}
We would like to thank to Samuel Wilkinson for his scientific reviews.

\bibliography{neurips_2019}

\newpage
\section*{Supplementary Materials}
\section*{A Case Study}
We present an example of a designer's workflow using our forward and reverse network to design and optimise urban volumes that would produce desired wind flow. A fictional site layout is visualised. This demo's bounding box's width and depth is 256 metres and maximum height is 64 meters. This is twice as large as our training set and shows the advantage of using a CNN.

Our neural network code is implemented using Tensorflow 2.0 and its Keras module. Communication between our CAD software and Tensorflow is by http requests responded by a Flask server. In our current setup, the bottleneck is pre-processing geometry, that needs to be voxelised. This can be sped up in the future using external mesh libraries.

\subsection*{A.1 Initial Sketch of Volumes}
Initial sketches of urban volumes can be drawn in a CAD software to provide a desired idea of the design and produce initial CFD analysis. Optionally this step can be omitted and a designer can directly draw a point-cloud representing areas with slow wind (step A.3).

\subsection*{A.2 Initial Interactive CFD Analysis}
\begin{figure}[h]
  \centering
  \includegraphics[width=\textwidth]{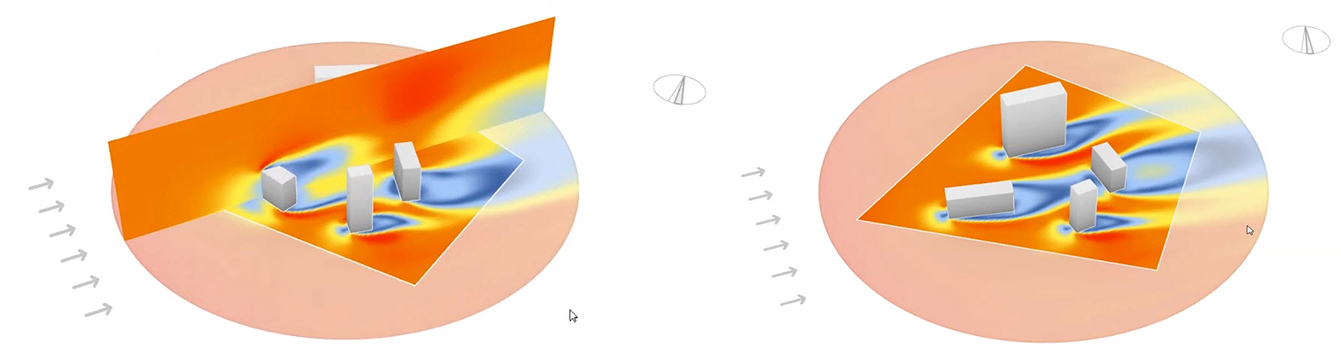}
  \caption{Initial CFD.}
\end{figure}
Our forward trained network provides a spatial CFD analysis prediction within a few seconds and is visualised in our CAD software.

\subsection*{A.3 Thresholded and Modified CFD Analysis}
\begin{figure}[h]
  \centering
  \includegraphics[width=\textwidth]{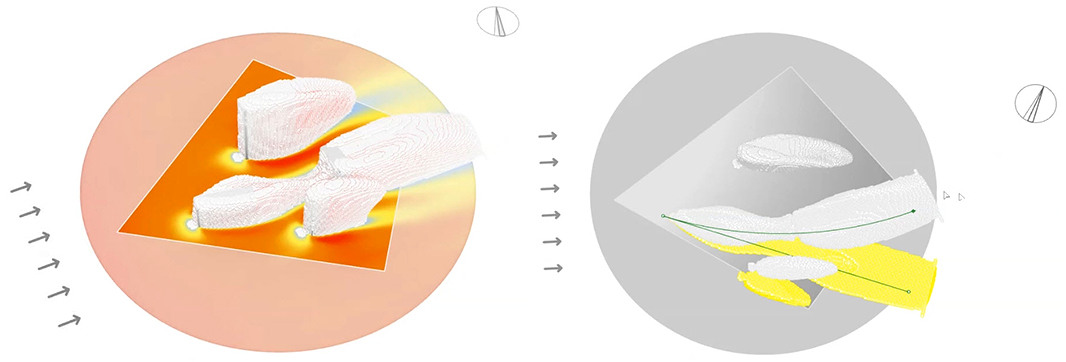}
  \caption{Threshold.}
\end{figure}
The CFD is thresholded to localise only areas with low wind speeds. These are areas better-suited for outdoor activity like sitting and running.
A point-cloud represents these areas, which can be freely modified using geometry transformations to design desired wind effect.

\subsection*{A.4 Geometry Prediction}
\begin{figure}[h]
  \centering
  \includegraphics[width=\textwidth]{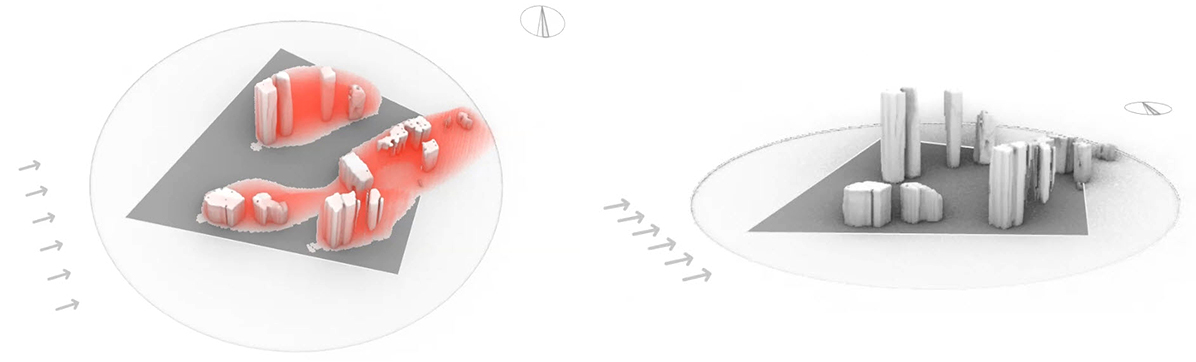}
  \caption{Geometry prediction.}
\end{figure}
Our reverse trained network predicts urban volumes that will produce our target wind flow and can be exported as a mesh object.

\subsection*{A.5 Final CFD Analysis}
\begin{figure}[ht]
  \centering
  \includegraphics[width=\textwidth]{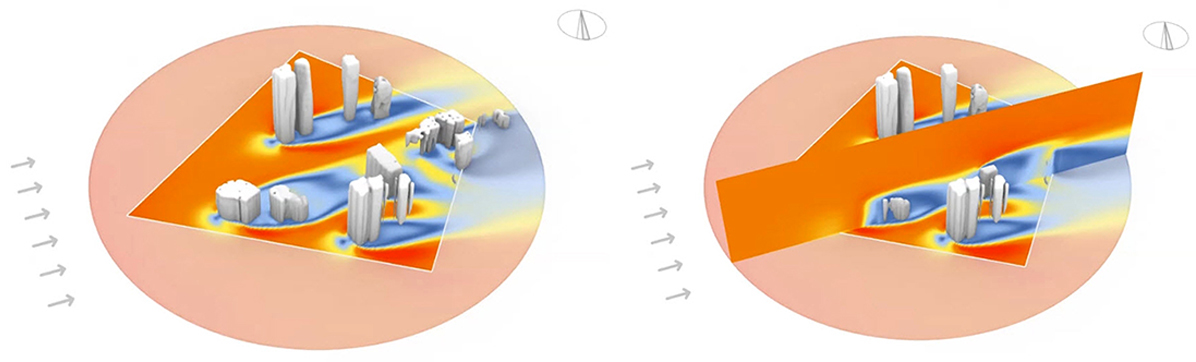}
  \caption{Final CFD.}
\end{figure}
Predicted volumes are used to do a CFD prediction of the wind flow. 

\subsection*{A.6 Discussion}
In future research we will look at interior spaces, in which case passive cooling plays a significant role in minimising energy consumption. Input for reverse direction would improve if pedestrian comfort was used for example.

Our current limitation is that only one direction of wind is considered. This is applicable in areas where there is a dominant wind direction. In other areas, multiple directions (usually 36) must be considered. Our forward trained network can predict all those directions and will be combined then.

\end{document}